\documentstyle[11pt,aaspp4,psfig]{article}

\def\etal{{\it et al.\ }}
\def\kms{km~s$^{-1}$}

\def\msun{{\rm M}_{\sun}}

\begin{document}

\hskip 3.4truein To appear in {\it The Astronomical Journal}

\vskip 0.5truein

\title{Neutral Gas Distribution and Kinematics of the \\
Nearly Face--on Spiral Galaxy NGC 1232}
\author{Liese van Zee\footnote{Jansky Fellow}}
\affil{National Radio Astronomy Observatory,\footnote{The National Radio 
Astronomy Observatory is a facility of the National Science Foundation,
operated under a cooperative agreement by Associated Universities Inc.}
PO Box 0, Socorro, NM 87801}
\affil{lvanzee@nrao.edu}
\authoremail{lvanzee@nrao.edu}
\author{Jessica Bryant}
\affil{Agnes Scott College, Decatur, GA 30030}
\affil{jbryant@agnesscott.edu}
\begin{abstract}
We have analyzed high velocity
resolution HI synthesis observations of the nearly 
face--on Sc galaxy NGC 1232.  The neutral gas distribution 
extends well beyond the optical extent of the galaxy.
As expected, local peaks in the HI column density are 
associated with the spiral arms.  Further, the HI column
density drops precipitously near the center of the galaxy.
Closed contours in the velocity field suggest either that 
the system is warped, or that the rotation curve declines.  
The velocity dispersion is 
approximately constant throughout the system, with a median 
value of 9.9 $\pm$ 1.8 \kms.  When corrected for rotational broadening, 
there is no indication of a radial trend in 
the neutral gas velocity dispersion in this galaxy.
\end{abstract}


\section{Introduction}

NGC 1232 is a nearly face--on, gas--rich, Sc galaxy.  Face--on
galaxies provide excellent targets for studies of radial
properties, such as the chemical enrichment of the interstellar
medium (e.g., McCall, Rybski, \& Shields \markcite{MRS85}1985;
 Vila--Costas \& Edmunds \markcite{VE92}1992;
 Zaritsky, Kennicutt, \& Huchra \markcite{ZKH94}1994;
Ferguson, Gallagher, \& Wyse \markcite{FGW98}1998) or the
velocity dispersion of the neutral gas (e.g., van der Kruit \& Shostak 
\markcite{vKS82}1982, \markcite{vKS84}1984; 
Shostak \& van der Kruit \markcite{SvK84}1984;
Dickey, Hanson, \& Helou \markcite{DHH90}1990;
Boulanger \& Viallefond \markcite{BV92}1992;
Kamphuis \& Briggs \markcite{KB92}1992;
Rownd, Dickey, \& Helou \markcite{RDH94}1994).
NGC 1232 first came to our attention as part
of a study of radial trends in elemental abundances 
(van Zee, Salzer, \& Haynes \markcite{vZ98a}1998a; 
van Zee \etal \markcite{vZ98b}1998b).
The outermost \ion{H}{2} region observed by van Zee \etal 
\markcite{vZ98a}\markcite{vZ98b}(1998a,b) has an
abnormally low oxygen abundance and a higher N/O ratio than
expected at that radius.  Such features may arise from  
asymmetries in the disk, or from perturbations induced by tidal
interactions.   We thus decided to investigate the
neutral gas distribution and kinematics of NGC 1232 to determine
if there were kinematic peculiarities associated with
this particular \ion{H}{2} region, or more generally
in the outer gas disk.

With a relatively low inclination angle, approximately
30\arcdeg, NGC 1232 is also a prime candidate for
studies of the neutral gas velocity dispersion as a function of
radius.  Since the HI distribution typically extends a factor 
of 1.3 to 2 times the size of the optical disk in Sc galaxies 
(e.g., Broeils \& Rhee \markcite{BR97}1997), one might
naively expect a decrease in the observed velocity dispersion
beyond the radius of active star formation
[star formation activity may not be truncated in spiral galaxies,
however (Ferguson \etal \markcite{FWGH98}1998b)]. 
Previous studies of the neutral gas velocity dispersion in face--on
spiral galaxies have been inconclusive.  While some observations
indicate that the velocity dispersion may decrease with radius
(e.g., NGC 6946, Boulanger \& Viallefond \markcite{BV92}1992),
others indicate that the velocity dispersion is constant in the outer
gas disk (e.g., NGC 1058, Dickey \etal \markcite{DHH90}1990). 
Such studies are hampered by two main problems: (1) rotational
broadening and (2) insufficient spectral resolution.  The
observations of NGC 1232 presented in this paper have
adequate spectral resolution, but suffer from significant
rotational broadening, in part due to a severe warp in
the gas disk.  Nonetheless, these observations  do provide
additional insight on  radial trends of the
neutral gas velocity dispersion in spiral galaxies. 

The physical parameters of NGC 1232 are summarized in 
Table \ref{tab:global}.  
Throughout this paper, we adopt a distance of 21.5 Mpc
(van Zee \etal \markcite{vZ98b}1998b), based on an
assumed H$_0$ of 75 \kms~Mpc$^{-1}$ and a Virgocentric
infall model. Optically, NGC 1232 is a rather
nondescript Sc galaxy with well defined spiral arms
and optical colors (de Vaucouleurs \etal \markcite{RC3}1991)
that are typical for late--type spiral galaxies.
Its total HI mass and M$_{\rm HI}$/L$_{\rm B}$ ratio
are also quite typical of the class. 
On the other hand,  it's far infrared
luminosity (Soifer \etal \markcite{SBNS89}1989) 
is significantly lower than typical for Sc galaxies 
(see, e.g., Roberts \& Haynes \markcite{RH94}1994 for
trends as a function of Hubble type).  

This paper is organized as follows.  The HI synthesis
observations and data reduction are presented in Section
\ref{sec:obs}.  The neutral gas distribution and kinematics
of NGC 1232 are discussed in Section \ref{sec:ana}.
A brief discussion and summary are presented
in Sections \ref{sec:disc} and \ref{sec:conc}. 
Finally, the results of optical spectroscopy
of the spatially nearby system NGC 1232A are summarized
 in an appendix.

\section{Observations and Data Reduction}
\label{sec:obs}
HI synthesis imaging observations of NGC 1232 were conducted by 
the NRAO summer students with the DnC configuration
of the Very Large Array\footnote{The Very Large Array
is a facility of the National Radio Astronomy Observatory.} (VLA)
on 1992 July 4.  The observations spanned 5 hours, with a 
total on--source integration time of 227 minutes.   
The correlator was used in 2AD mode with the
right and left circular polarizations tuned to 1684 \kms.  The total
band width was 1.56 MHz.  The on--line Hanning smoothing option
was selected, producing final spectral data cubes of 127 channels,
each 2.6 \kms~wide.  Standard tasks in AIPS
(Napier \etal \markcite{AIPS}1983)
were employed for calibration and preliminary data reduction.
Calibration of this data set was complicated by the
fact that the west arm of the array was shadowed during the
observations of the primary flux density calibrator, 3C 286.
However, the observations of 3C 286 were sufficient to set the flux
density scale; phase and bandpass calibration was
derived from observations of a nearby continuum source, B0237--233.
With an observed flux density of 6.21 Jy at 1412.5 MHz, the observations
of B0237--233 had adequate signal--to--noise for calibration of  
both the phase and the bandpass.

After calibration, the line data was transformed to the {\it x--y}~plane.
A robust weighting technique was employed by the AIPS task {\small \rm IMAGR}
to optimize the beam shape and noise levels of the final data cubes 
(Briggs \markcite{B95}1995).  The ``robustness parameter'' in 
{\small \rm IMAGR} controls the
weighting of the {\it u--v}~data, permitting a fine--tuning between sensitivity
and resolution.  As currently implemented, a robustness
of 5 corresponds to natural weighting of the {\it u--v}~data 
(maximizing sensitivity) while a robustness of --5 corresponds to
uniform weighting (lower sensitivity, but better spatial resolution).
The relevant {\small \rm IMAGR} parameters for three data cubes 
are listed in Table \ref{tab:maps}.  Throughout
this paper we will refer to the robustness of
1.0 cube as the ``low resolution cube,'' to the robustness of
0.5 cube as the ``intermediate weight cube,'' and to the
robustness of --0.5 cube as the ``high resolution cube.''
All subsequent analysis of the data cubes was performed within
the GIPSY package (van der Hulst \etal \markcite{GIPSY}1992).

Line emission from NGC 1232 filled most of the bandpass (channels 7 to 121).
Since the few line--free channels were at the extreme edge of the 
bandpass, they had poor noise characteristics and thus
were insufficient to create an accurate model of the
continuum emission in either the {\it u--v}~or {\it x--y}~planes. 
Continuum subtraction was a necessary step in the data reduction
process, however, since NGC 1232 has 58.5 mJy of continuum
emission at L--band (Condon \markcite{C87}1987).
Continuum subtraction was conducted in the {\it x--y}~plane,
with a continuum image created separately for each data cube.
To create the continuum image, the mean of the inner 100 channels 
(1553.8 $< v <$ 1814.3 \kms) was computed.  This image was severely 
contaminated by the line emission of the galaxy;
the region of galaxy emission was blotted from the image and
replaced with the value 0.
Next, the mean was computed for channels with velocities between 1538.2 and 
1561.6 \kms~(line emission from only the eastern half of the galaxy)
and for channels with velocities between 1811.7 and 1835.2 \kms~(line
emission from only the western half of the galaxy).
  The continuum sources coincident with the
galaxy emission were recovered from these two maps.  The final continuum
image for each resolution was inspected to verify that all of the continuum
sources were recovered; an example of one of these continuum images
(from the intermediate weight data cube) is shown in Figure
\ref{fig:cont}.    

The continuum image for each resolution was subtracted from every 
channel of the data cube to produce the final continuum--free data cubes. 
Selected channels from the continuum subtracted intermediate weight
data cube are presented in Figure \ref{fig:chans}.   The
continuum subtraction appears to have been successful
(see, for instance, the baseline of the integrated flux density
profile, Figure \ref{fig:flux}).  Note, however, that even if the continuum
subtraction had not been as successful, the subsequent analysis
of the gas kinematics and dynamics would still be valid.  
In particular, the velocity field
and velocity dispersion measurements are fairly robust, even in the
presence of continuum emission; only
the total gas distribution would have been significantly
effected by poor continuum subtraction.  

To determine if the total HI flux density was recovered in the
HI synthesis observations, the flux density was measured in each channel
of the continuum subtracted data cubes.  
The flux density profile, corrected for primary beam attenuation,
 is shown in Figure \ref{fig:flux}.  Qualitatively, the
flux density profile is similar to those obtained from single dish observations
(Fisher \& Tully \markcite{FT81}1981; Reif \etal \markcite{R82}1982;  
Staveley--Smith \& Davies \markcite{SSD88}1988).
The total flux density recovered 
from the low resolution data cube was 122.4 $\pm$ 12.2 Jy \kms, in good 
agreement with previous single dish measurements (e.g., 115.5 $\pm$ 7.2 
Jy \kms, Green Bank 43m [Fisher \& Tully 
\markcite{FT81}1981]; 127.9 $\pm$ 13.8 Jy \kms, 
Jodrell Bank [Staveley--Smith \& Davies \markcite{SSD88}1988]).  Thus,
despite the calibration problems and missing short spacings, the flux density
scale appears to be correct, and the total HI flux density has been recovered.

The slight asymmetry between the two horns of the integrated HI profile
 has been ascribed to  emission
from the spatially nearby galaxy NGC 1232A (ESO 547--G016) (e.g., Reif \etal 
\markcite{R82}1982; Becker \etal \markcite{BMRvW88}1988). However, there is no 
apparent gas density enhancement or perturbation in the velocity field near 
this object (see Figure \ref{fig:mom}).  Furthermore,
a heliocentric optical velocity of $6570. \pm 15.$ \kms~was measured
from several emission lines with the Palomar 5m telescope;\footnote{Observations 
at the Palomar Observatory were made as part of a continuing cooperative agreement
between Cornell University and the California Institute 
of Technology.} further details of the emission lines observed in
this galaxy are presented in the appendix.
This newly derived redshift is in basic agreement 
with the velocity listed in NED;\footnote{The NASA/IPAC 
Extragalactic Database (NED) is operated by the Jet Propulsion Laboratory,
California Institute of Technology, under contract with the National
Aeronautics and Space Administration.} thus, this object does
not appear to be physically associated with NGC 1232.

Both moment analysis and Gaussian fitting were used to characterize the
gas distribution, velocity field, and velocity dispersion of NGC 1232.  
Moment maps of each data cube were computed in the following manner.
First, the low resolution cube was smoothed to a resolution of twice 
the beam; second, the smoothed cube was clipped at the 2$\sigma$ level; the resultant
clipped cube was then interactively blanked to remove spurious noise spikes.
Signal was identified based on spatial continuity between channels.   A
conditional transfer was used to blank the corresponding locations in 
the low, intermediate, and high resolution data cubes (corrected for primary
beam attenuation) based on the blanked, smoothed
cube.  Moment maps of the blanked data cubes were created with the GIPSY task 
{\small \rm MOMENTS}.  In addition,  Gaussian fits to the observed line 
profiles at each position in the unclipped data
cubes were obtained with the GIPSY task {\small \rm GAUFIT}.  In regions
of high signal--to--noise,  Gaussian fitting is preferred over moment
analysis since it is less susceptible to beam smearing.   The
resultant maps were qualitatively similar to the moment maps in
both gas distribution and velocity field.  However, significant
differences were found in the velocity dispersion maps, probably
due to beam smearing.  Thus, the moment maps were used
for investigation of the gas distribution and kinematics, while interactive 
Gaussian fitting was employed for study of the velocity dispersion
as a function of position.  The zeroth and first moments are
presented in Figure \ref{fig:mom}; the optical images used in this
figure are described more fully in van Zee \etal \markcite{vZ98b}(1998b).

\section{Analysis}
\label{sec:ana}

\subsection{Gas Distribution and Velocity Field}
\label{sec:mom0}

There are several notable features in the gas distribution of NGC 1232. 
First, the HI distribution extends well beyond the optical
radius.  At a column density limit of 
10$^{20}$ atoms cm$^{-2}$, the HI--to--optical diameter ratio is
1.6, quite typical of spiral galaxies
(e.g., Cayatte \etal \markcite{CKBvG94}1994; 
Broeils \& Rhee \markcite{BR97}1997).  
Second, as has been seen in many other massive
galaxies (e.g., Brinks \& Bajaja \markcite{BB86}1986; 
 Broeils \& van Woerden \markcite{BvW94}1994),
the neutral gas column density is lower in the center of the galaxy.
The azimuthally averaged radial profile of the gas density is
shown in Figure \ref{fig:rot}d and tabulated in Table \ref{tab:rot}.
As seen in Figure \ref{fig:rot}d, the eastern and western
sides of the galaxy are not perfectly symmetric.  The eastern half of the
galaxy has a higher column density, corresponding to the prominent spiral
arms on that side.  On the western side, the gas distribution extends
to larger radii, perhaps due to a warp of the gas disk
in the outer regions.  The observed asymmetry between the horns of the
global HI profile is the result of the convolution of this asymmetric 
gas distribution with the rotation of the disk.  Such asymmetries are not 
expected to be long lived, but also are not uncommon in spiral galaxies 
(e.g., Haynes \etal \markcite{H98}1998).

Closed contours in the velocity field (Figure \ref{fig:mom}d) 
suggest that the outer gas disk is warped, causing an apparent
decline in the rotation curve.   A model for the rotation
curve of NGC 1232 was derived using tilted--ring analysis of the intermediate
resolution velocity field.   The process of fitting a rotation curve
is iterative; we began by fitting both sides of the velocity field
in order to determine the kinematic center and systemic velocity 
(Table \ref{tab:global}).  Next, the center and systemic velocity 
were held fixed and each side was fit separately to determine the position and 
inclination angles.   The position angle was well constrained by the models.
However, the fit to the inclination angle was relatively unconstrained.
Based on the optical axial ratios, the inclination angle was initially assumed
to be approximately 30\arcdeg.  Since the {\small \rm ROTCUR} fits resulted
in wild variations of the inclination angle, the inclination was set to 30\arcdeg~in
the inner regions of the galaxy and then permitted to slowly decrease in the
outer regions.  Finally, the rotation curve for each side was determined by fixing 
the inclination and position angles to the average of both sides with the
additional constraint that these values must change smoothly from ring to ring.
The result of {\small \rm ROTCUR}'s fit to the approaching and receding sides are 
shown in Figure \ref{fig:rot}; Table \ref{tab:rot} lists the average values
for both sides.  As seen in the upper panel of Figure \ref{fig:rot}, a
warped disk model is able to recover a flat rotation curve in the outer gas disk.  
However, since the inclination angle was poorly constrained by the tilted--ring
models, we cannot rule out the possibility that the inclination angle is
constant at all radii (i.e., the gas disk may not be warped).  
Thus, a rotation curve was also derived using the same parameters as the warped
disk model, with the exception that the inclination angle was held fixed at
30\arcdeg~throughout the disk; not unexpectedly, this model results in
a falling rotation curve in the outer disk. 

Figure \ref{fig:pv} shows the position--velocity diagram for a slice
along the major axis of NGC 1232.  The rotation curve derived from
the warped disk model is superposed.  As frequently found in spiral
galaxies, the warp appears to begin at the edge of the optical disk (see e.g.,
NGC 4013, Sparke \& Casertano \markcite{SC88}1988; NGC 4138, Jore, 
Broeils, \& Haynes \markcite{JBH96}1996).  

Both the warped disk and falling rotation curve models adequately
reproduce the major features of the velocity field.  However,
several kinks in the iso--velocity contours are found in the outer 
regions of the gas disk which cannot be well fit by either of these
models.  Such kinks are usually associated with spiral streaming motions, 
but these features occur in regions well beyond the prominent spiral arms.
Note, however, that such features would be difficult to detect
in the inner regions due to the presence of a steeply rising rotation curve.  
Thus, their apparent absence in the inner region of the galaxy does not 
eliminate the possibility that they arise from the continuation of spiral 
density waves to the edge of the disk.   Due the the steep rotational 
gradient, higher spatial resolution observations will be needed 
to investigate fully the pattern speeds in the outer gas disk. 
 
\subsection{Rotation Curve Analysis}
\label{sec:mom1}

The derived rotation curves were used to determine the dark matter content
of NGC 1232.  Rotation curve decomposition requires knowledge of the gaseous
and stellar mass distributions (assumed to be disk--like) and an assumed dark matter
distribution.  The contribution of the gaseous and stellar components
to the gravitational potential were calculated using the GIPSY routine
{\small \rm ROTMOD}.  The input surface densities were estimated from a series 
of ellipses with position angles of 270\arcdeg~and inclinations of 30\arcdeg. 
The stellar distribution was taken as the observed azimuthal average
to the extent of the optical emission (Figure \ref{fig:rot}e), followed by an 
extrapolated exponential disk.  In fitting the rotation curve, the HI distribution 
was multiplied by 4/3 to account for the He content.  The gas content was held 
fixed during the fitting process.  Note that for the observed neutral gas 
distribution of NGC 1232, the net acceleration is outwards for the inner
part of the gas disk; following the usual convention, this outward acceleration 
is denoted by a negative velocity.
The stellar disk was fit with one free parameter,
the mass--to--light ratio (M/L$_B$).  Note that since the available R--band
image was obtained under non--photometric conditions, the observed surface
brightness distribution was scaled to match the total blue luminosity 
(de Vaucouleurs \etal \markcite{RC3}1991).
While B--band and R--band scale lengths are typically slightly different for spiral galaxies
(see, e.g., de Jong \markcite{d96}1996), this minor difference should be insignificant for 
the present mass models.  A simple spherical halo model was used
for the dark matter contribution; this model has
only two free parameters, the core radius, R$_c$, and the central density, $\rho_o$:
\begin{equation}
\rho(r) = \rho_o \left(1 + \left({r \over {\rm R}_c}\right)^2\right)^{-1},
\end{equation}
where $\rho$ is the dark matter density at radius $r$.
  In practice, the maximum velocity of the halo 
is fit rather than the central density:
\begin{equation}
{\rm V}^2_{halo}(r) = {\rm V}^2_{H} \left[1 - {{\rm R}_c \over r} \arctan \left({r \over 
{\rm R}_c}\right)\right],
\end{equation}
where V$_{halo}$ is the contribution of the halo to the gravitational
potential at a radius $r$ and 
V$_{H}$ is the asymptotic velocity, equal to $\sqrt{4\pi {\rm G} \rho_o {\rm R}^2_c}$.

To determine the parameters of the dark matter halo, the derived rotation 
curves were fit using a maximum--disk method.  In this 
method, the disk mass--to--light ratio is fit using the inner part of the rotation 
curve without overshooting the derived circular velocities.  Initially, the maximum
mass--to--light ratio (in the B--band) is found by assuming no dark matter is present.  
For the warped disk model, the rotation curve derived from only the observable mass 
rapidly underestimated the observed rotation curve at large radii.  Thus, the
presence of a dark matter halo was assumed and the dark matter halo core radius,
maximum velocity, and disk mass--to--light ratio were fit in an iterative manner.  
In contrast, the falling rotation curve model was well fit by the luminous
matter alone and did not require a dark matter halo (i.e., the falling
rotation curve is nearly Keplarian in nature).  The derived rotation
curve decompositions are shown in Figure \ref{fig:rotcur}.  In both cases,
the derived disk mass--to--light ratios (3.0 for a warped disk, 3.7 for
a falling rotation curve) are reasonable for the observed colors (see, e.g., 
Broeils \markcite{B92}1992).  

One significant difference between the two models is the derived dynamical mass. 
For a spherical mass distribution, the dynamical mass within radius $r$ is:
\begin{equation}
M_T(r) = 2.326 \times 10^5~V^2(r)~r
\end{equation}
where $V(r)$ is in km s$^{-1}$ and $r$ is in kpc.
However, if no dark matter halo is required, the system is disk--like and the 
dynamical mass is significantly less than that calculated assuming a spherical halo.
For the falling rotation curve model, the dynamical mass is simply the sum of the 
gaseous and stellar mass, $1.83 \times 10^{11}~\msun$.  For the warped
disk model, the dark matter halo dominates at large radii, so the dynamical
mass is given by the above equation, $4.36 \times 10^{11}~\msun$~at the
last measured point of the rotation curve.  This latter value
gives a dark--to--luminous
ratio of 1.9, quite typical of Sc galaxies (Broeils \markcite{B92}1992).

\subsection{Line Shapes and Velocity Dispersion}
\label{sec:mom2}

As mentioned in Section \ref{sec:obs}, interactive Gaussian
fitting was used to determine the velocity dispersion throughout the galaxy.
Both the second moment map and the velocity dispersion map
from {\small \rm GAUFIT} showed an apparent trend in the
velocity dispersion as a function of radius, with the inner
regions of the galaxy having extremely high velocity dispersions.
Despite the relatively low inclination of this galaxy, however, the velocity
dispersion measurements are significantly contaminated by
rotational broadening throughout the galaxy.  For instance,
for beams between 40 and 120 arcsec from the center,
the average rotational gradient is 0.3 \kms~arcsec$^{-1}$; for beams between
210 and 300 arcsec, the average rotational gradient is --0.2 \kms~arcsec$^{-1}$.
With a 40\arcsec~beam, this corresponds to
a minimum of 8 \kms~of rotational broadening per beam.  Only in
the outer regions of the galaxy is the velocity gradient sufficiently
shallow to permit detailed analysis of the line shape and width.
A line profile from gas in the outer disk is shown in Figure \ref{fig:prof}.  
This profile was created by averaging line profiles from regions on the
eastern and western sides of the galaxy with radii between 200\arcsec~and 260\arcsec.
The profile is well fit by a Gaussian with a velocity dispersion of 10.0~\kms.

Despite the large rotational gradient, it is possible to investigate 
radial trends in the velocity dispersion by deconvolving the rotational 
and turbulent motions.  The contribution from rotational broadening can be
calculated either from a rotation curve model (e.g., Table \ref{tab:rot}), or
from the observed rotational gradient, as measured in the position--velocity 
diagram.  The latter method is more direct, but is limited to regions along 
the major axis.  However, the major axis is also the region where the 
rotational broadening is minimized; therefore, in the following analysis
we limited our study to two independent slices $\pm$24\arcsec~from the major
axis of the high resolution data cube.  The velocity dispersion was
measured interactively for each beam averaged region and
the rotational gradient for each region was obtained 
directly from the position--velocity diagram.  Figure \ref{fig:disp} shows 
the results of the deconvolution of thermal and rotational broadening; each slice
has been folded so that all radii are positive.  The observed line widths were 
corrected for rotational broadening by assuming that the rotational and thermal 
components add in quadrature.  As illustrated in Figure \ref{fig:disp}, 
the velocity dispersion is approximately constant as a function of radius, 
with an average value of 9.9 $\pm$ 1.8 \kms.   This is in remarkably good agreement
with the velocity dispersion measured from the average of profiles
in the outer regions of the galaxy.    

Similar results were also obtained from more detailed rotational models.
An estimate of the rotational broadening throughout the
disk was obtained by creating a model system with the GIPSY task
{\small \rm GALMOD}; the system was assumed to have a negligible
intrinsic velocity dispersion and the observed
gas distribution and the derived (no warp) rotation
curve were given as the input parameters.   The derived velocity 
dispersion of this model system is
thus indicative of the rotational broadening at any given position in
the galaxy.  The intrinsic velocity dispersion was recovered by
subtracting, in quadrature, the model dispersion from the
observed velocity dispersion of the high resolution data cube
(pixel--by--pixel).  Similar to the above method, no radial trend was 
seen in the final velocity dispersion map.  The average velocity 
dispersion was measured to be 10.3 $\pm$ 1.7 \kms~throughout the gas disk
using this technique.

\section{Discussion}
\label{sec:disc}

 The observed velocity dispersion is much higher than
one might expect for thermal motions.  For thermal broadening
at 21 cm, the velocity dispersion is related to the thermal temperature by:
\begin{equation}
T~\sim~121~\sigma_v^2.
\end{equation}
Thus, a velocity dispersion of 9.9 \kms~corresponds to a Doppler temperature
of $\sim$11900 K, which is significantly higher than the expected temperature
of the warm neutral medium (c.f. Kulkarni \& Heiles \markcite{KH88}1988). 
However, similarly high velocity dispersions have
been observed in several other galaxies (e.g., NGC 3938, van der Kruit \& Shostak
\markcite{vS82}1982; NGC 1058, Dickey \etal \markcite{DHH90}1990; NGC 6946,
Boulanger \& Viallefond \markcite{VB92}1992; M101, Kamphuis \markcite{K93}1993).  
Since neither the previous nor the present observations have adequate spatial 
resolution to resolve individual HI clouds, the observed velocity dispersion is
the ensemble average of several clouds per beam.   In fact, the mean motions of 
the gas clouds, rather than the temperature of the gas within each cloud, could
be the dominant motion resulting in the broad velocity dispersion.

For a galaxy with a high star formation rate, turbulent motion is
obviously a significant contributor to the velocity dispersion
in the inner regions of the galaxy, where supernovae and outflows provide 
large amounts of kinetic energy.  One difficulty, however, is to explain
how such high levels of turbulence are sustained throughout the gas disk,
particularly in regions well beyond the optical edge or in systems with 
modest star formation rates (such as NGC 1232).  
Recently, Sellwood \& Balbus \markcite{SB99}(1999) proposed that
MHD instabilities in the differentially rotating gas disk could
support a moderate amount of turbulence at all radii via dynamical
heating of the disk.  Their model requires a modest magnetic field
($\sim$ 3 $\mu$G) to reproduce a velocity dispersion of 6 \kms~in the
outer gas disks of spiral galaxies.  While it is still unclear
whether this mechanism is the dominant source of turbulence at large radii,
it is an intriguing possibility.

Several studies report a radial decrease in the
velocity dispersion, with the outer regions of a galaxy having
narrower and more perfectly Gaussian line shapes (e.g., 
NGC 6946, Boulanger \& Viallefond \markcite{VB92}1992; 
M101, Kamphuis \markcite{K93}1993).
As discussed previously, NGC 1232 also has an observable radial
decrease in the velocity dispersion, all of which can be accounted
for by rotational broadening in the presence of a steeply rising rotation
curve.  While most studies acknowledge the complications of
rotational broadening, it can be quite difficult to account fully for its
effects on the data.  For instance, when convolved with a decreasing
gas density in the center of the galaxy, beam smearing can result in
highly asymmetric, non--Gaussian spectra.  A velocity dispersion 
measured either by moment analysis or Gaussian fitting
will not accurately reflect the width of the line in such spectra.
Thus, it is extremely difficult to study the profile shape
and radial trends in the velocity dispersion
 unless the observations are obtained with  high
spectral {\it and} spatial resolution. 

\section{Conclusions}
\label{sec:conc}

We have investigated the gas distribution and kinematics of the
nearly face--on spiral galaxy NGC 1232.  Our results and
conclusions are summarized below.

(1) As is typical of spiral galaxies, the neutral gas distribution
extends a factor of 1.6 beyond the optical radius.  The gas 
distribution is slightly asymmetric, with a higher gas column
density associated with the prominent spiral arms on the eastern side.

(2) The observed rotation curve has a nearly Keplarian decline
beyond the edge of the optical disk.  Small variations in the
inclination angle (which is poorly constrained due to the low
inclination of this galaxy) can recover a flat rotation curve
for the system.  Assuming that the system is indeed warped,
the dark--to--luminous ratio is 1.9.

(3) The presence of a steeply rising rotation curve complicates
measurements of the velocity dispersion as a function of radius. 
Nonetheless, after correcting for rotational broadening, the velocity
dispersion appears to be constant throughout the galaxy,
with an average value of 9.9 $\pm$ 1.8 \kms.

As mentioned in the introduction, one of the reasons for looking at
the gas distribution of NGC 1232 was to see if there were unusual 
gas kinematics associated with the outermost HII regions.  
There are no obviously disturbed kinematic features in the present
data set, but such features may have been washed out by 
the poor spatial resolution of the observations.  Nonetheless,
to first order, the spectra are all well fit by a single Gaussian
(with the exception of the central region where the rotational
broadening is severe) and are consistent with a slightly 
warped, differentially rotating, gas disk extending well beyond
the optical radius.  Thus, it is unlikely that large scale
kinematic peculiarities are the cause of the abnormal abundances
measured in the outer HII region.

Analysis of this data set emphasizes the need for good quality high
resolution (both spectral {\it and} spatial) observations of face--on
galaxies.  The few canonical systems (e.g., NGC 1058 and NGC 3938) have
been studied extensively with both WSRT and the VLA, but more systems
are needed to address possible dependences of the velocity dispersion on galaxy
morphology and star formation rates.   While the requirement for low inclination 
angles significantly restricts the potential target list, corrections for rotational
broadening are less important with high spatial resolution observations.  In fact,
these studies will become more feasible within the next few years; with 
a new correlator, the upgraded VLA will permit high velocity resolution 
over a wide bandpass, thereby improving the quality of similar observations
 of moderately inclined spiral galaxies.

\acknowledgements
We thank the NRAO summer students of 1992 who elected to observe
NGC 1232 as part of their summer student project: Chris De Pree, Robbie
Dohm--Palmer, Eric Olson, Ben Oppenheimer, Carlos Rabaca, James
Sweeney, and Eric Wagoner.  We thank Michael Rupen, Katrina Koski, 
Andreea Petric, and Dave Westpfahl for many informative conversations 
about gas in face--on galaxies.  Martha Haynes kindly provided
access to the Palomar 5m telescope for the redshift determination
of N1232A.  We thank the anonymous referee for thoughtful comments on
rotational broadening.  Robbie Dohm--Palmer and Katrina Koski are gratefully
acknowledged for providing helpful comments on an early draft of the paper.
JB acknowledges funding through the Research Experience for 
Undergraduates program run by the NRAO and funded by the National 
Science Foundation.

\appendix
\section{Metallicity of NGC 1232A}

As noted in Section \ref{sec:obs}, optical spectra of NGC 1232A were 
obtained with the Palomar 5m telescope during the night of 1999 January 23.  
The spectra were of sufficient quality that further analysis (beyond the
redshift determination) was warranted.  Here, we describe the observations
in more detail and present the results of  oxygen and nitrogen
abundance calculations for the two HII regions observed.  

During the observations, a long slit (2\arcmin) with a 2\arcsec~aperture  was
centered on NGC 1232A for one 1200 sec exposure.
The slit was positioned at an angle of 0\arcdeg~(the parallactic
angle at the time of the observation) and passed through both
the center of the galaxy and an HII region 19\arcsec~to the
south  (Figure \ref{fig:n1232a}).  A 5500 \AA~dichroic was used to 
split the light to the two 
sides (blue and red), providing complete spectral 
coverage from 3600--7600 \AA.  The blue spectra were acquired
with the 600 lines/mm diffraction grating (blazed at 4000 \AA).
The red spectra were acquired with the 316 lines/mm diffraction
grating (blazed at 7500 \AA).  Thinned 1024$\times$1024 Tek CCDs,
with read noises of 8.6 e$^-$ (blue) and 7.5 e$^-$ (red), were
used on the two sides of the spectrograph.  Both CCDs had
a gain of 2. e$^-$/DN.  The effective spectral resolution of
the blue camera was 5.0 \AA~(1.72 \AA/pix); the effective spectral
resolution of the red camera was 7.9 \AA~(2.47 \AA/pix).
The spatial scale of the long slit was 0.62 \arcsec/pix on the
blue and 0.48 \arcsec/pix on the red side.

	While the night was non--photometric, relative flux calibration
was obtained by observations of standard stars from the list
of Oke \markcite{Oke90}(1990).  Wavelength calibration was obtained by 
observations of arc lamps taken before and after the galaxy observations.
A Hollow Cathode (Fe and Ar) lamp was used to calibrate the
blue spectra; a combination of He, Ne, and Ar lamps were used
to calibrate the red spectra.

	The spectra were reduced and analyzed with the IRAF\footnote{IRAF 
is distributed by the National Optical Astronomy Observatories.} package.
The spectral reduction included bias subtraction, scattered light
corrections, and flat fielding with both twilight and dome flats.
The 2--dimensional images were rectified based on the arc lamp
observations and the trace of stars at different positions along 
the slit.  The sky background was removed from the 2-dimensional
images by fitting a low order polynomial along each row of the
spectra.  One dimensional spectra of the two HII regions 
were extracted from the rectified images using a 3.6\arcsec~extraction 
region for the southern HII region and a 3.0\arcsec~region for
the central HII region (Figure \ref{fig:n1232a}).  Reddening corrected 
line strengths for both HII regions are listed in Table \ref{tab:lines}.

Derivation of the oxygen abundance for both HII regions
followed the methods described in van Zee \etal \markcite{vZ98b}(1998b).
The temperature sensitive line [OIII] $\lambda$4363 was detected
in the southern HII region and thus the electron temperature 
could be determined directly; for this HII region, the oxygen
abundance was calculated using the emissivity coefficients from
a version of the FIVEL program (De Robertis \etal \markcite{DDH87}1987)
assuming a low density (100 cm$^{-3}$) and a
 T$_{\rm e}$ of 11050 $\pm ^{2530}_{1210}$.  
For the central HII region, the
oxygen abundance was estimated based on the strong line ratios
(R$_{\rm 23} \equiv$ ([OII] + [OIII])/H$\beta$) using the
model grid of McGaugh \markcite{M91}(1991).  An
electron temperature of 6500 $\pm$ 500 K
reproduced this estimated oxygen abundance, and was thus adopted
to derive the N/O ratio for the HII region.
The derived oxygen and nitrogen
abundances for both HII regions are listed in Table \ref{tab:lines}.
Not unexpectedly, the southern HII region is relatively low abundance
while the central HII region is significantly higher abundance.
With measurements of only two HII regions in hand, the derivation
of an abundance gradient in this system is premature; nonetheless,
to first order, the abundance gradient is similar to those found
in other spiral galaxies (e.g., Vila--Costas \& Edmunds \markcite{VE92}1992; 
Zaritsky \etal \markcite{ZKH94}1994; van Zee \etal \markcite{vZ98b}1998b).


\begin{deluxetable}{ll}
\tablewidth{30pc}
\tablecaption{Physical Properties of NGC 1232 \label{tab:global}}
\tablehead{
\colhead{Parameter} & \colhead{Value}  } 
\startdata
Distance (Mpc)                         & 21.5\tablenotemark{a}\nl
Kinematic Center (B1950)               & 03 07 30.2, --20 46 08 \nl
\nl
HI Properties: \nl
\qquad Integrated {H \small \rm I} flux density (Jy km s$^{-1}$)   & 122.4 $\pm$ 12.2 \nl
\qquad Heliocentric velocity (km s$^{-1}$)             & 1678.9 $\pm$ 1.0 \nl
\qquad {H \small \rm I} profile width (50\%) (km s$^{-1}$)          & 234 $\pm$ 2.  \nl
\qquad {H \small \rm I} profile width (20\%) (km s$^{-1}$)          & 256 $\pm$ 3.  \nl
\qquad Velocity Dispersion  (km s$^{-1}$)              & 9.9 $\pm$ 1.8 \nl
\qquad {H \small \rm I} size at 10$^{20}$ atoms cm$^{-2}$ (arcmin $\times$ arcmin) & 11.6 $\times$ 9.7 \nl
\qquad Peak {H \small \rm I} surface density (atoms cm$^{-2}$)      & 1.4 $\times$ 10$^{21}$ \nl
\nl
Optical Properties: \nl
\qquad Apparent blue magnitude         &  10.52 $\pm$ 0.14\tablenotemark{b} \nl
\qquad (U--B)$_0$                      & --0.02 $\pm$ 0.10\tablenotemark{b} \nl
\qquad (B--V)$_0$                      &   0.60 $\pm$ 0.04\tablenotemark{b} \nl
\qquad R$_{\rm 25}$ (kpc)              &  23.2\tablenotemark{b} \nl
\qquad R$_{\rm d}$ (kpc)               &  6.3\tablenotemark{a} \nl
\nl
Derived Properties: \nl
\qquad Hydrogen mass, $M_{\rm HI}$ (10$^{10}\,M_{\sun}$)    & 1.33 \nl
\qquad Optical luminosity, $L_{\rm B}$ (10$^{10}\,L_{\sun}$)& 4.46 \nl
\qquad FIR luminosity, $L_{\rm FIR}$ (10$^{10}\,L_{\sun}$)  & 0.875\tablenotemark{c} \nl
\qquad Dynamical mass, $M_{\rm T}$ (10$^{10}\,M_{\sun}$)    & 43.6 \nl
\qquad Dynamical mass -- no warp, $M_{\rm T}$ (10$^{10}\,M_{\sun}$)    & 18.3 \nl
\qquad $M_{\rm HI}/L_{\rm B}$ ($M_{\sun}/L_{\sun}$)         & 0.3 \nl
\qquad $M_{\rm HI}/M_{\rm T}$                               & 0.03 \nl
\qquad $M_{\rm dark}/M_{\rm lum}$                           & 1.9 \nl
\enddata
\tablenotetext{a}{van Zee et al.\ 1998}
\tablenotetext{b}{de Vaucouleurs et al.\ 1991}
\tablenotetext{c}{Soifer et al.\ 1989}
\end{deluxetable}

\begin{deluxetable}{ccccrcc}
\tablewidth{30pc}
\tablecaption{Parameters of the HI Data Cubes \label{tab:maps}}
\tablehead{
\colhead{} &\colhead{} & \colhead{}& \colhead{linear} \\[.2ex]
\colhead{Robustness}& \colhead{Beam}& \colhead{rms}  & \colhead{resolution} \\[.2ex]
\colhead{Parameter} & \colhead{[arcsec$\times$arcsec]} & \colhead{[mJy beam$^{-1}$]} & \colhead{[kpc beam$^{-1}$]} } 
\startdata
  1   & 53.9~$\times$~~26.2 & 1.7 & 5.6 $\times$ 2.7 \nl
  0.5 & 50.3~$\times$~~24.0 & 1.7 & 5.2 $\times$ 2.5 \nl
--0.5 & 43.2~$\times$~~21.1 & 2.0 & 4.5 $\times$ 2.2 \nl
\enddata
\end{deluxetable}

\begin{deluxetable}{rcccc}
\tablewidth{30pc}
\tablecaption{Rotation Curve of NGC 1232 \label{tab:rot}}
\tablehead{
\colhead{radius} & \colhead{$\Sigma_{\rm HI}$} & \colhead{i} & \colhead{PA}&\colhead{V$_c$} \\[.2ex]
\colhead{[arcsec]}& \colhead{[M$_{\odot}$ pc$^{-2}$]} & \colhead{[\arcdeg]}& \colhead{[\arcdeg]}& \colhead{[km s$^{-1}$]} } 
\startdata
10.  & 1.27 & 30 & 270 &  70.7 $\pm$ 20.7   \nl
40.  & 2.20 & 30 & 270 & 184.9 $\pm$  1.4   \nl
70.  & 4.23 & 30 & 270 & 209.9 $\pm$  1.6   \nl
100. & 5.85 & 30 & 270 & 216.6 $\pm$  0.8   \nl
130. & 6.34 & 30 & 270 & 221.8 $\pm$  0.7   \nl
160. & 5.91 & 30 & 270 & 220.1 $\pm$  0.6   \nl
190. & 5.14 & 29 & 270 & 221.0 $\pm$  0.6   \nl
220. & 3.70 & 28 & 269 & 220.9 $\pm$  0.7   \nl
250. & 2.18 & 26 & 268 & 219.4 $\pm$  1.3   \nl
280. & 1.19 & 23 & 268 & 218.8 $\pm$  1.7   \nl
310. & 0.72 & 21 & 268 & 215.8 $\pm$  1.6   \nl
340. & 0.52 & 20 & 268 & 219.1 $\pm$  1.9   \nl
370. & 0.41 & 20 & 269 & 217.9 $\pm$  2.3   \nl
400. & 0.32 & 20 & 269 & 217.8 $\pm$  6.4   \nl
\enddata
\end{deluxetable}

\begin{deluxetable}{lccc}
\tablewidth{35pc}
\tablecaption{Optical Line Intensities for NGC 1232A \label{tab:lines}}
\tablehead{
\colhead{Ionic} & \colhead{Rest} &\colhead{NGC 1232A (000+000)}& \colhead{NGC 1232A (000--019)}   \\[.2ex]
\colhead{Species} &\colhead{Wavelength} &\colhead{I($\lambda$)/I(H$\beta$)}&\colhead{I($\lambda$)/I(H$\beta$)}  }
\startdata
[OII]     &  3728 & 2.32 $\pm$ 0.17 & 2.30 $\pm$ 0.14   \nl
[NeIII]   &  3869 &      \nodata    & 0.19 $\pm$ 0.02   \nl
HI         & 4340 & 0.47 $\pm$ 0.05 & 0.47 $\pm$ 0.03   \nl
[OIII]    &  4363 &      \nodata    & 0.037 $\pm$ 0.016 \nl
HI        &  4861 & 1.00 $\pm$ 0.06 & 1.00 $\pm$ 0.05   \nl
[OIII]    &  4959 & 0.61 $\pm$ 0.05 & 1.34 $\pm$ 0.06   \nl
[OIII]    &  5007 & 1.81 $\pm$ 0.11 & 4.33 $\pm$ 0.19   \nl
HeI       &  5876 & 0.09 $\pm$ 0.01 & 0.10 $\pm$ 0.01   \nl
[OI]      &  6300 & 0.09 $\pm$ 0.12 & 0.064 $\pm$ 0.005 \nl
[OI]      &  6364 & 0.03 $\pm$ 0.01 & 0.020 $\pm$ 0.003 \nl
[NII]     &  6584 & 0.11 $\pm$ 0.01 & 0.045 $\pm$ 0.004 \nl
HI        &  6563 & 2.86 $\pm$ 0.22 & 2.84 $\pm$ 0.18   \nl
[NII]     &  6584 & 0.38 $\pm$ 0.03 & 0.14 $\pm$ 0.01   \nl
HeI       &  6678 &     \nodata     & 0.038 $\pm$ 0.004 \nl
[SII]     &  6716 & 0.53 $\pm$ 0.04 & 0.26 $\pm$ 0.02   \nl
[SII]     &  6731 & 0.27 $\pm$ 0.02 & 0.15 $\pm$ 0.01   \nl
[ArIII]   &  7136 &    \nodata      & 0.079 $\pm$ 0.006 \nl
\nl
\hline 
c$_{H\beta}$ &                & 0.02 $\pm$ 0.08 & 0.24 $\pm$ 0.06 \nl
F(H$\beta$)$\times10^{15}$ &  & 2.05 & 4.98 \nl
EW(H$\beta$) [\AA] &          & 10.8 & 95.3 \nl
log(R$_{\rm 23}$)&  & 0.676 $\pm$ 0.019 & 0.902 $\pm$ 0.013 \nl
T(O$^{++}$)  &                & 6500 $\pm$ 500 & 11050 $\pm ^{2530}_{1210}$  \nl
12 + log(O/H)  &              &  8.79 $\pm$ 0.10 & 8.21 $\pm$ 0.10 \nl
log(N/O) &                    & --1.24 $\pm$ 0.14 & --1.42 $\pm$ 0.15 \nl
\enddata
\end{deluxetable}

\psfig{figure=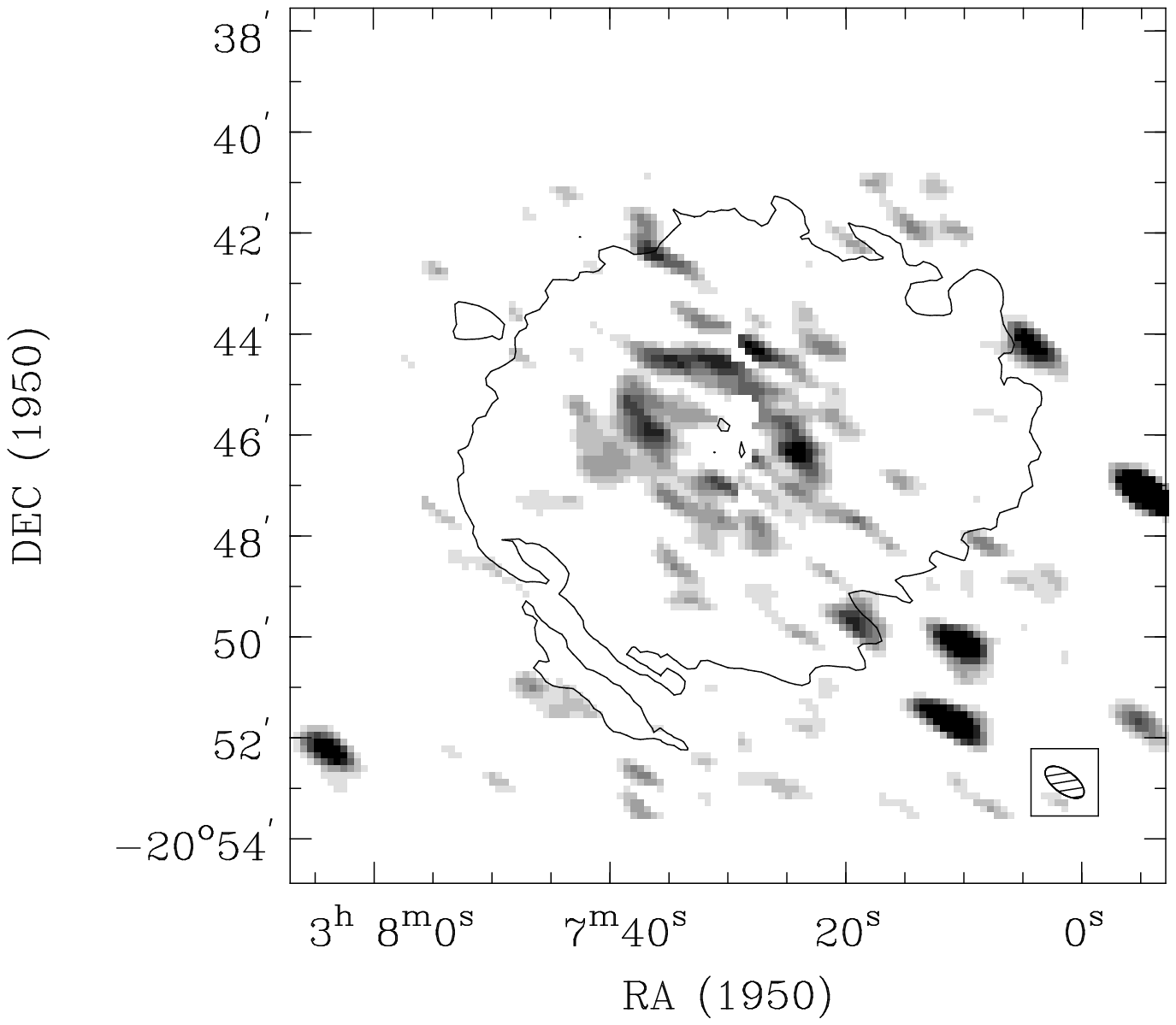,width=6.in,bbllx=50pt,bblly=20pt,bburx=500pt,bbury=600pt}
\vskip -0.3 truein
\figcaption[]{Continuum emission from the intermediate weight data cube. 
The beam size is 50.3$\times$24.0 arcsec. The 
region of HI emission is outlined at a column density 
limit of 10$^{20}$ atoms cm$^{-2}$.  The inner ring of continuum emission
is spatially coincident with the spiral arms.
\label{fig:cont} }

\psfig{figure=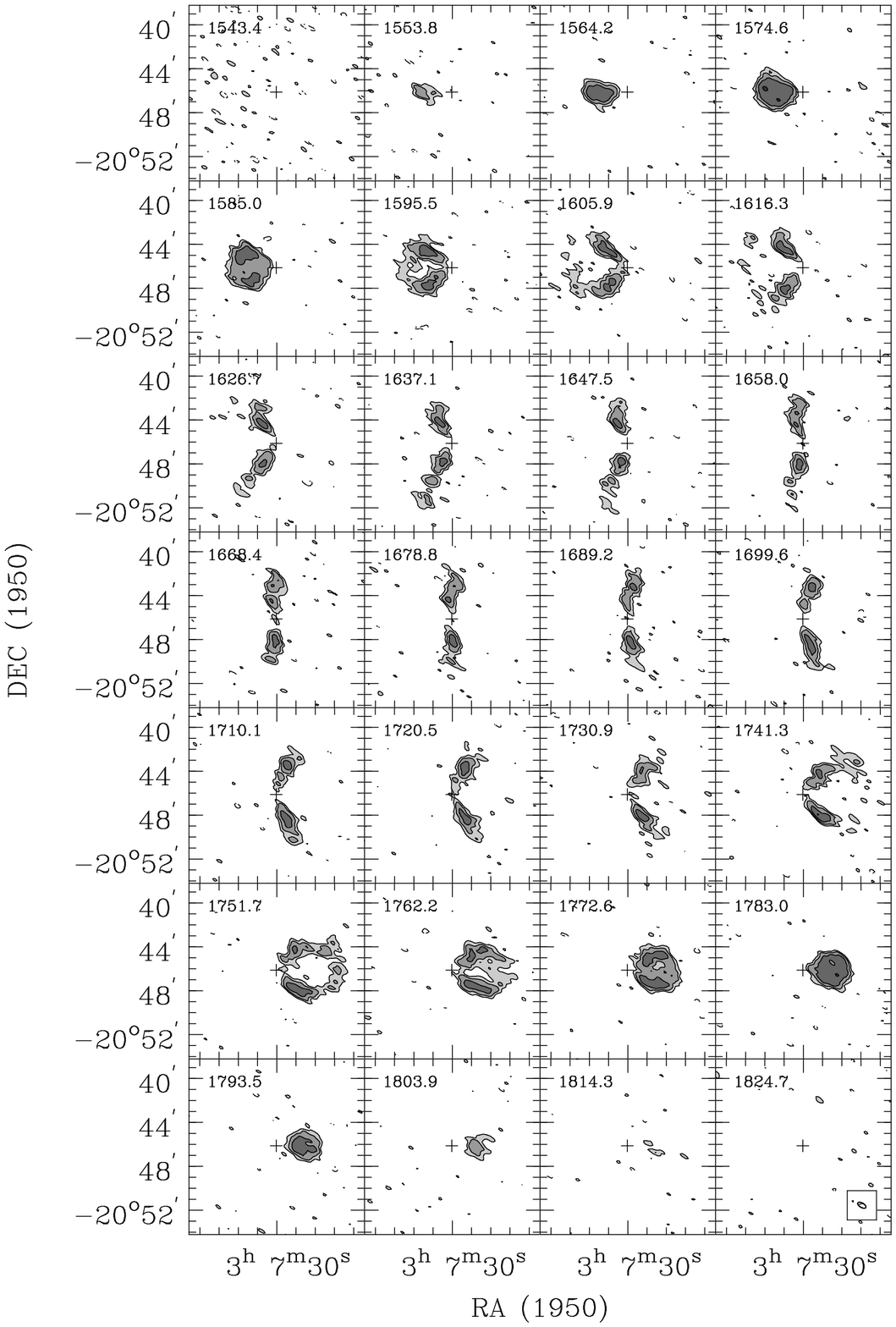,width=6.in,bbllx=50pt,bblly=40pt,bburx=550pt,bbury=650pt}
\vskip -0.3 truein
\figcaption[]{Selected channels from the intermediate weight HI data cube.  
Every fourth channel is shown.  The beam size, illustrated in the lower right panel,
 is 50.3$\times$24.0 arcsec. The contours represent
-3$\sigma$, 3$\sigma$, 6$\sigma$, and 12$\sigma$.  
The dynamical center is marked with a cross. \label{fig:chans} }

\psfig{figure=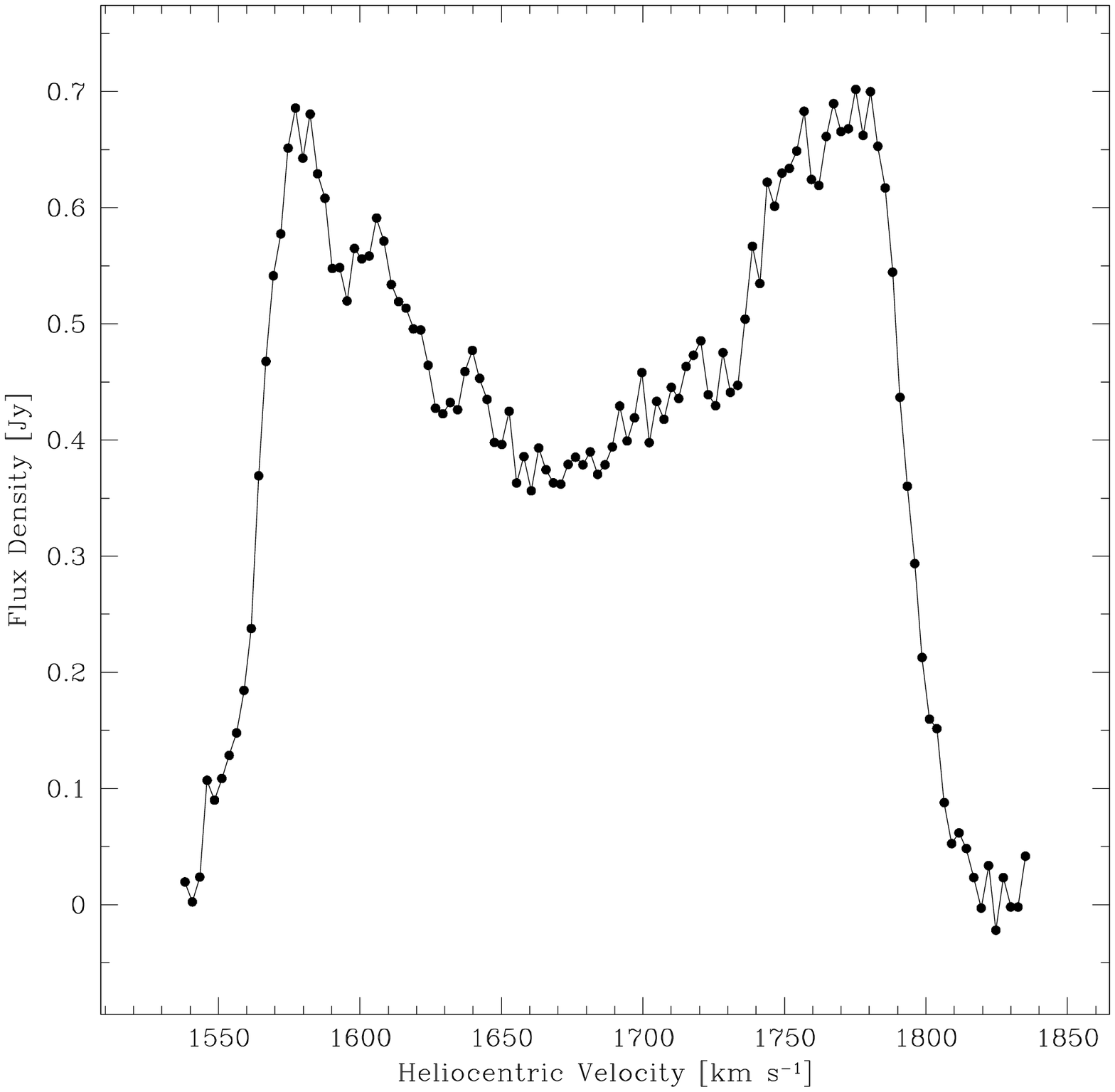,width=6.in,bbllx=0pt,bblly=40pt,bburx=600pt,bbury=700pt}
\vskip -1.3 truein
\figcaption[]{ The global HI profile for NGC 1232 derived from the continuum
subtracted low resolution HI data cube.  The total flux density is recovered in
the VLA observations. \label{fig:flux} } 

\centerline{\hbox{
\psfig{figure=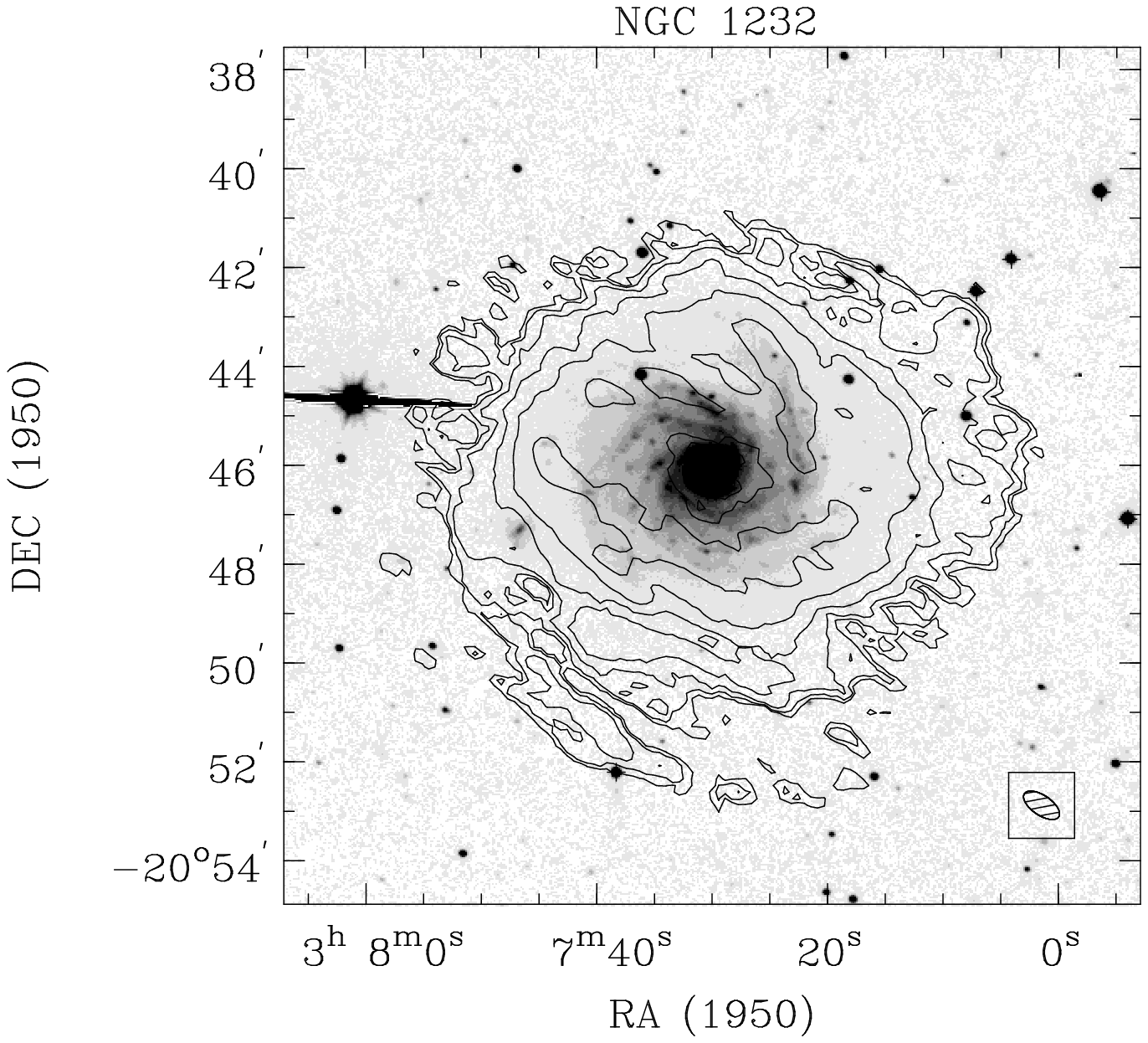,width=3.5in,bbllx=70pt,bblly=50pt,bburx=480 pt,bbury=398 pt,clip=t}
\psfig{figure=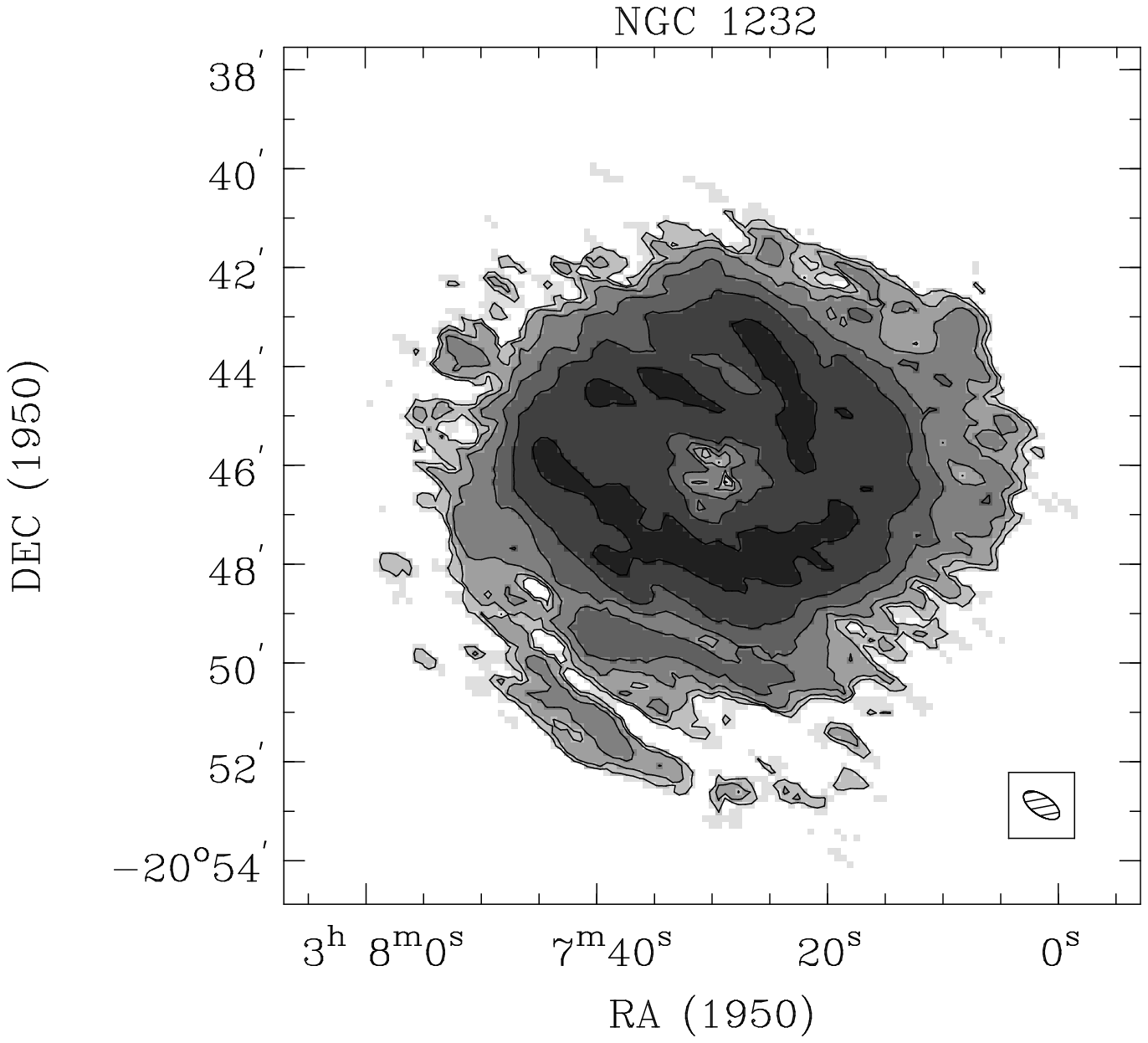,width=3.5in,bbllx=70pt,bblly=50pt,bburx=480 pt,bbury=398 pt,clip=t}
}}
\vskip 0.5truein

\centerline{\hbox{
\psfig{figure=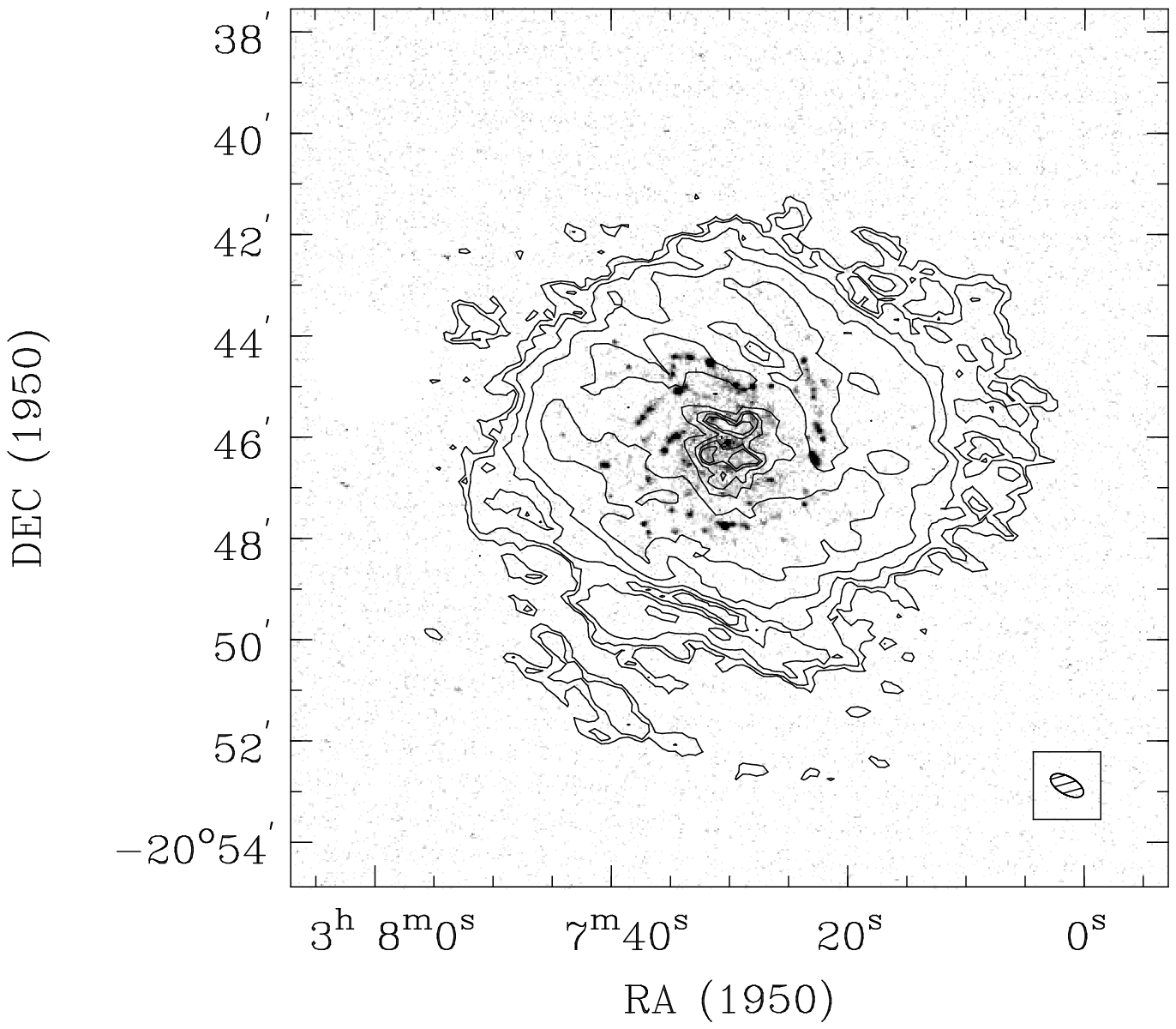,width=3.5in,bbllx=70pt,bblly=50pt,bburx=480 pt,bbury=398 pt,clip=t}
\psfig{figure=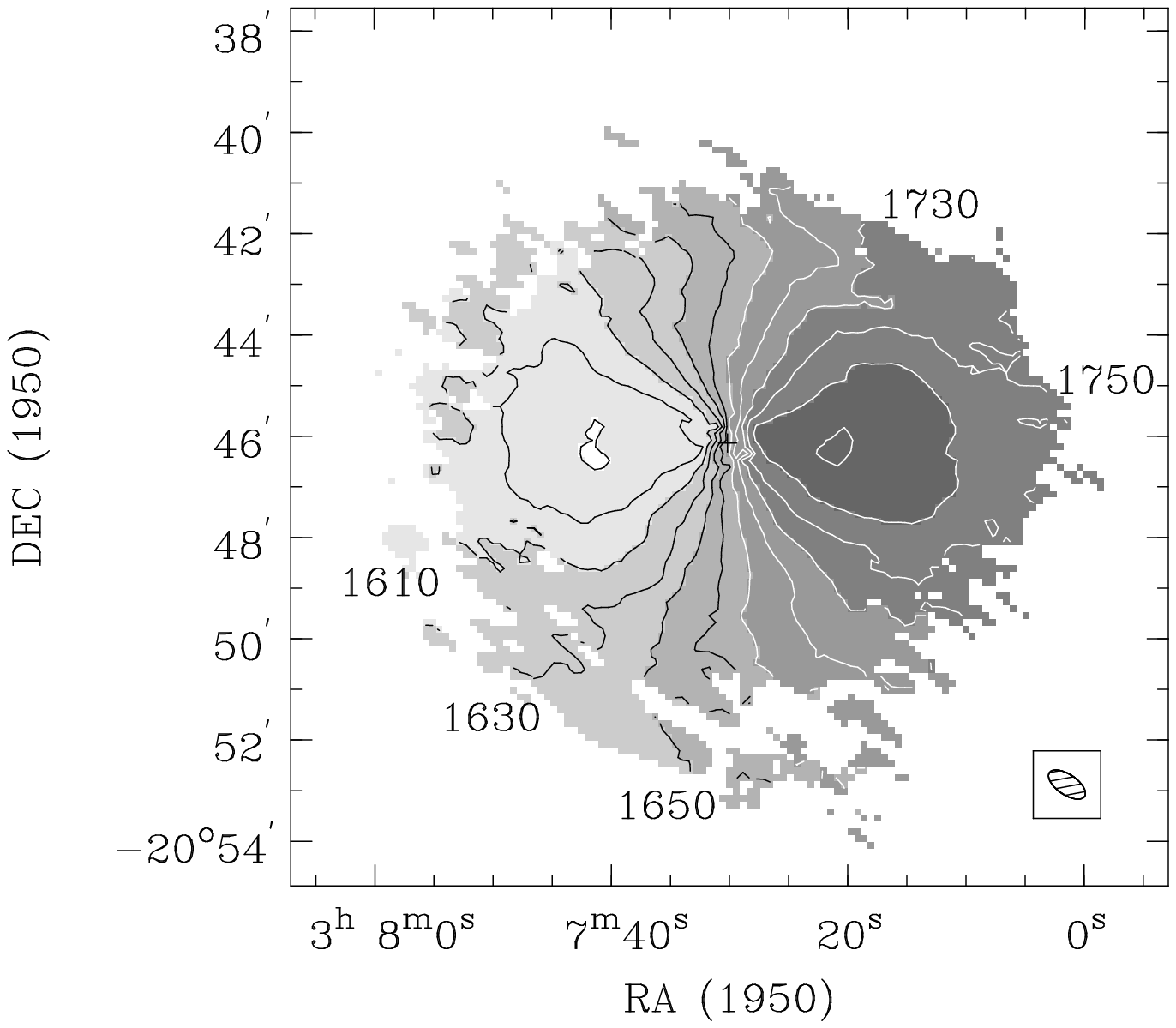,width=3.5in,bbllx=70pt,bblly=50pt,bburx=480 pt,bbury=398 pt,clip=t}
}}
\vskip 0.1truein

\figcaption[]{Moment maps of NGC 1232.  (a) The HI contours from the
intermediate weight data cube are shown overlaid on an optical R band image. 
The HI contours are 0.3, 0.6, 1.2,
2.4, 4.8, 9.6, and 19.2 $\times 10^{20}$ cm$^{-2}$.  The HI beam size is
50.4 $\times$ 24.0 arcsec.  The pixel scale of the 
optical image is 2.03 arcsec pixel$^{-1}$. 
(b) Same as (a) but with the HI column density distribution in both
grey scale and contours. 
(c) The HI contours from the high resolution data cube overlaid on
an H$\alpha$ image.  The HI contours are 0.5, 1, 2, 4, and 8 $\times$ 10$^{20}$ 
atoms cm$^{-2}$ with a beam size of 43.2 $\times$ 21.1.  The pixel scale of 
the H$\alpha$ image is 2.03 arcsec pixel$^{-1}$.
(d)  The velocity field of the intermediate weight data cube. 
The contours are marked every 20 \kms. \label{fig:mom} }

\psfig{figure=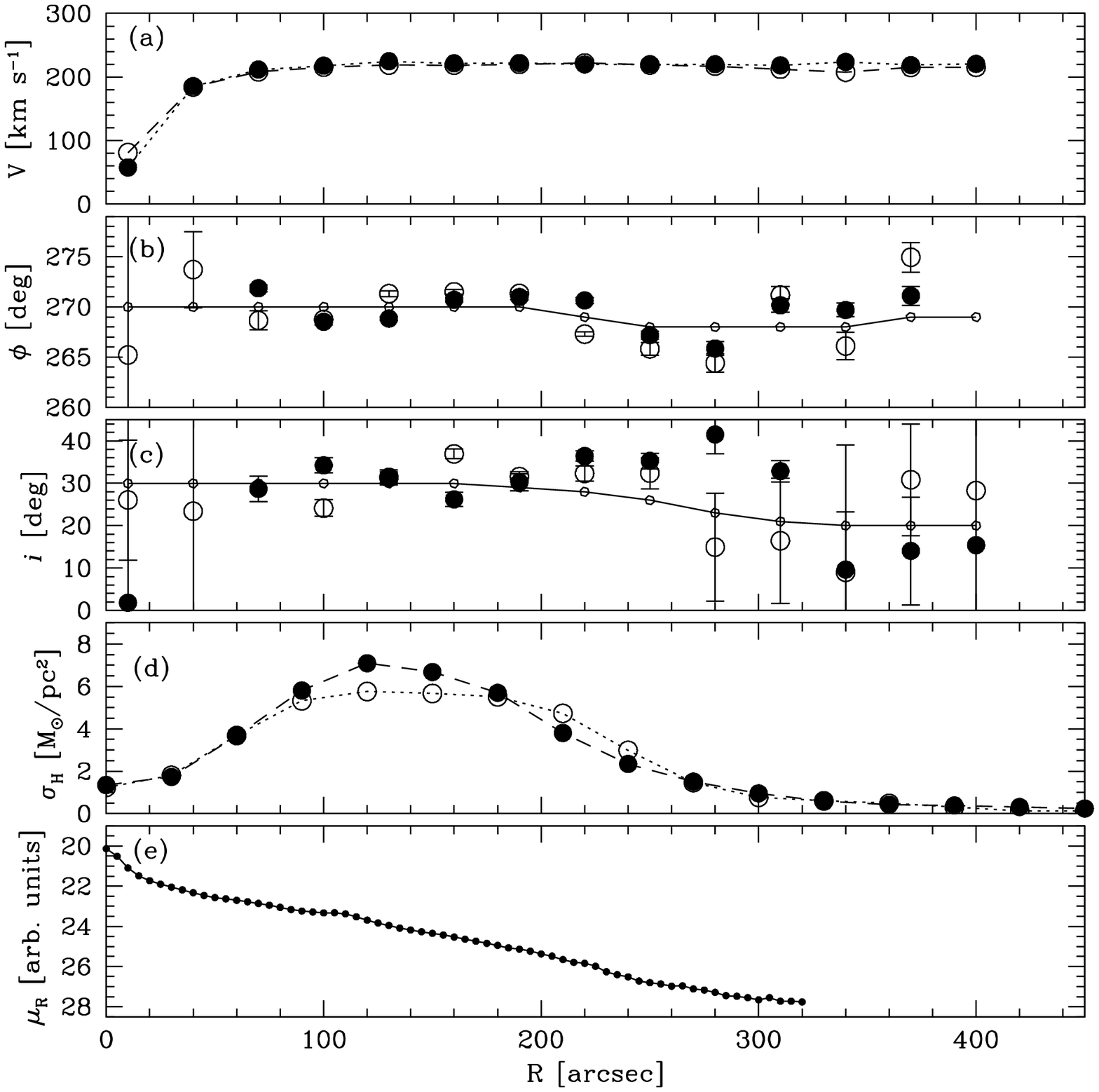,width=6.in,bbllx=0pt,bblly=100pt,bburx=550pt,bbury=700pt}
\vskip -0.5 truein
\figcaption[]{ Rotation curve fitting parameters: (a) the derived rotation 
curve, (b) position angle, (c) inclination angle, (d) neutral gas surface density,
and (e) optical surface brightness.  
The rotation curve was fit in GIPSY using an iterative process. 
The approaching (eastern) and receding (western) sides are represented
by filled and open circles, respectively. 
The values used to derive the rotation curve are indicated by the connected points.
The R--band optical surface brightness is displayed in arbitrary units since the
image was taken under non--photometric conditions.
\label{fig:rot} }

\psfig{figure=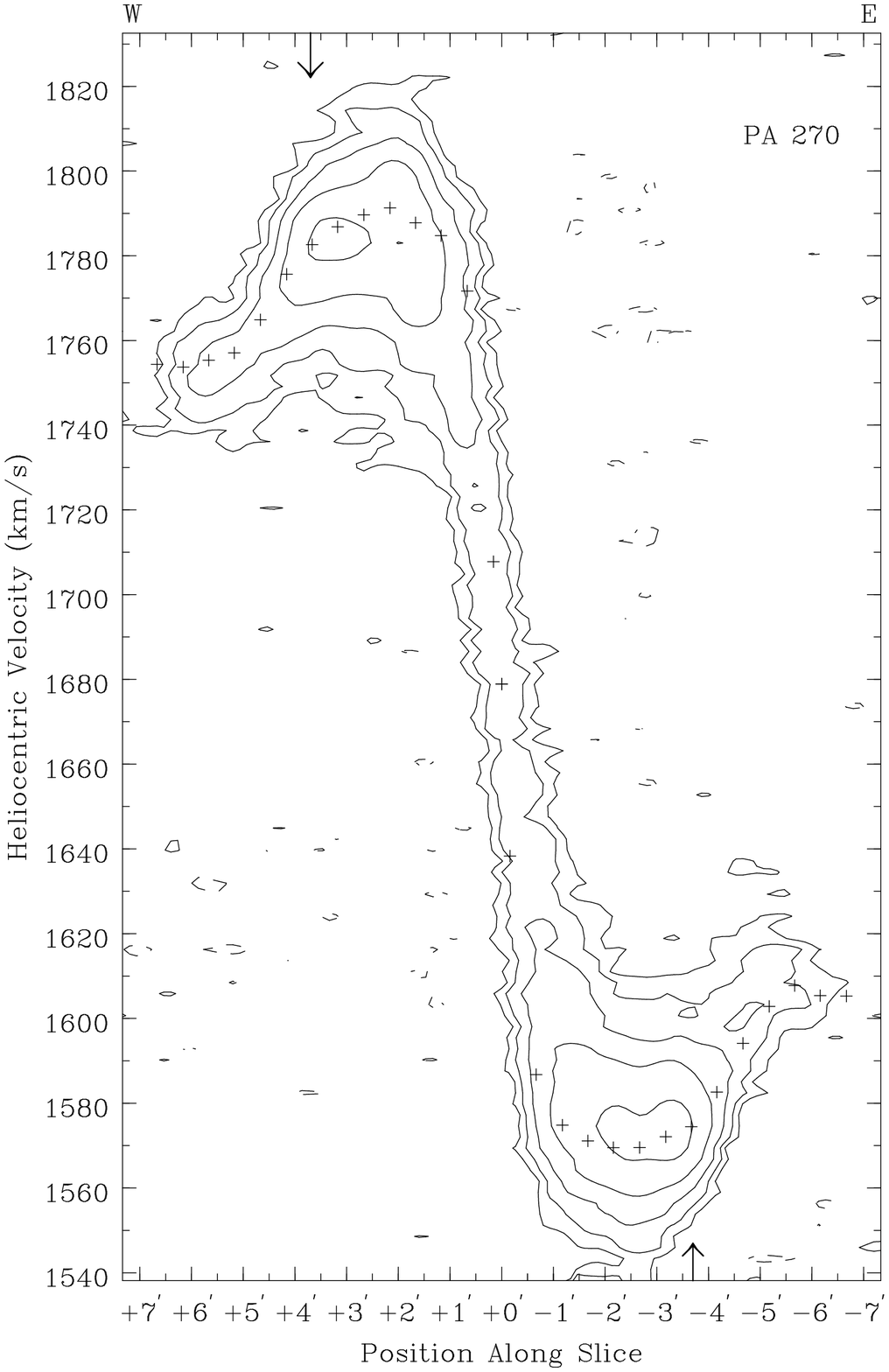,width=6.in,bbllx=50pt,bblly=30pt,bburx=650pt,bbury=800pt}
\vskip -0.3 truein
\figcaption[]{ A position-velocity diagram cut at a position angle of 270\arcdeg.
The derived rotation curve is marked by crosses.   Radii corresponding to R$_{25}$ 
are denoted by arrows.  The observed falling rotation
curve has been modeled here as a warp, although other interpretations
are possible. \label{fig:pv} }

\psfig{figure=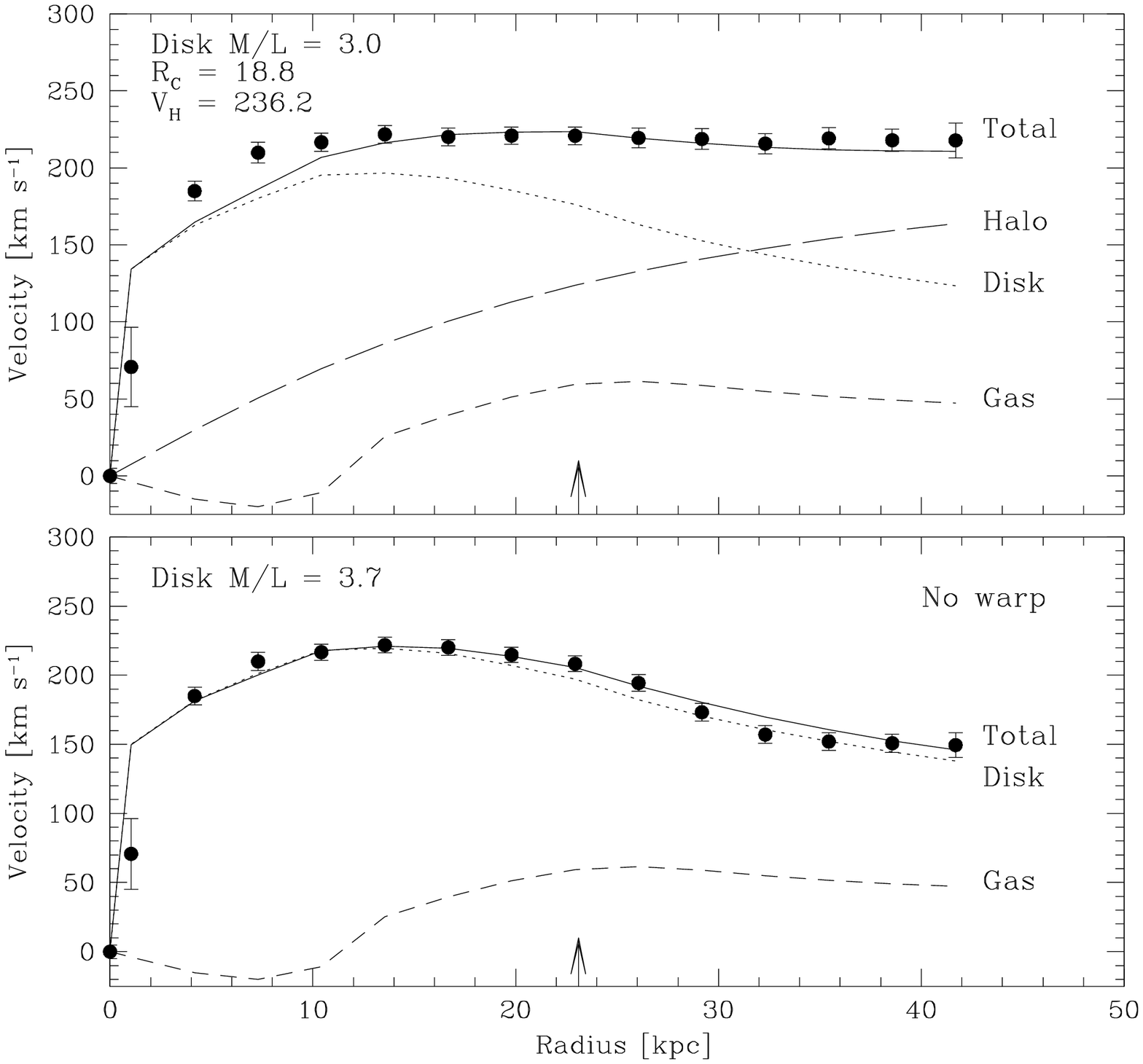,width=6.in,bbllx=20pt,bblly=100pt,bburx=550pt,bbury=750pt}
\vskip -0.6 truein
\figcaption[]{  Maximum disk rotation curve decomposition.  
The upper panel shows the rotation curve derived by using the parameters shown in Figure 5.  
The lower panel shows the rotation curve derived by assuming a constant inclination 
of 30\arcdeg.  In both these panels, the total rotation curve (solid line) is the 
quadrature sum of the mass contributions of the stellar and gaseous disks and the dark
matter halo.  Note that in the inner regions, the mass distribution of gaseous disk 
results a net acceleration outwards (denoted by a negative velocity in these plots).
The radius corresponding to R$_{25}$ is denoted by an arrow.
\label{fig:rotcur} }

\psfig{figure=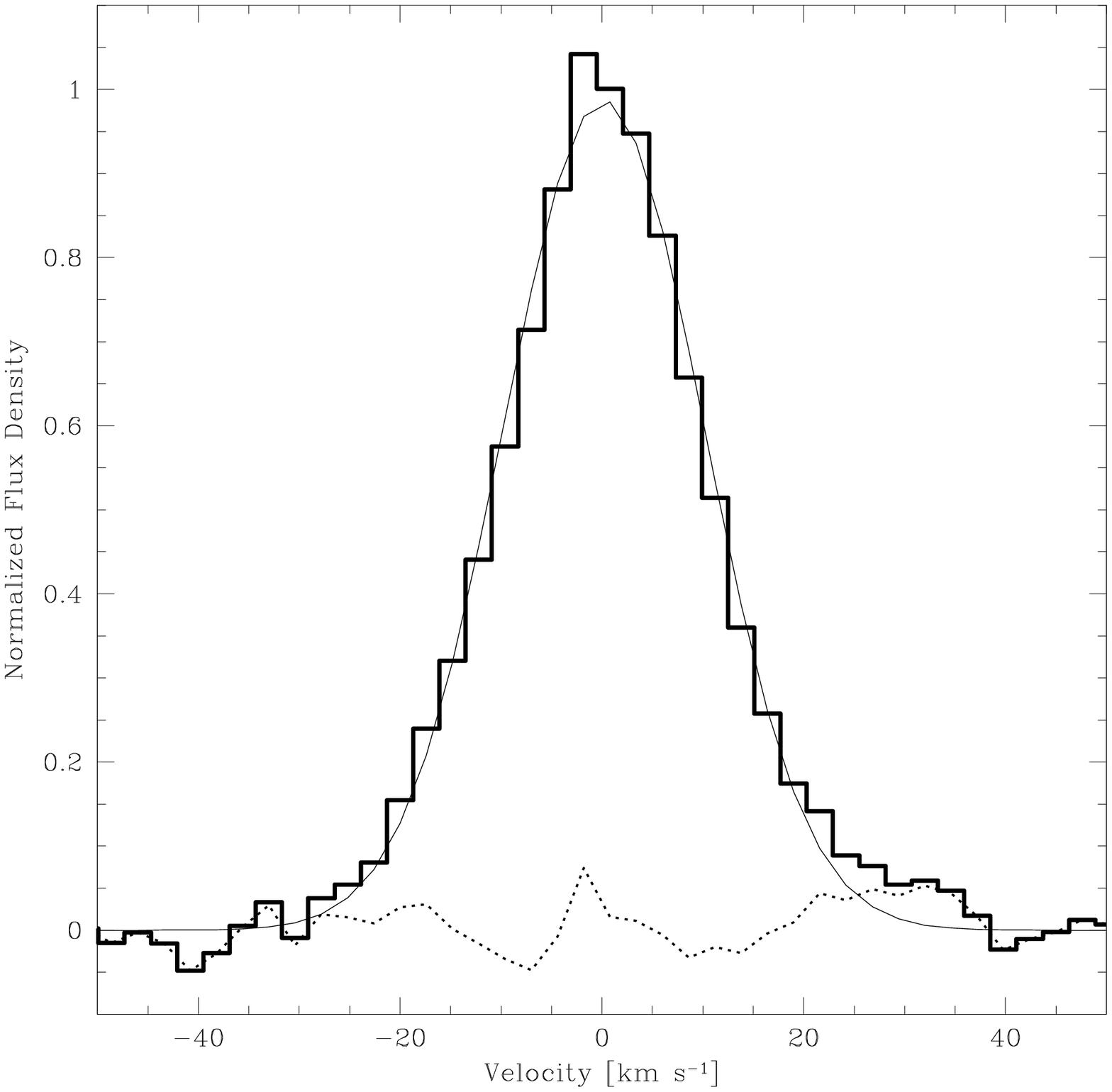,width=6.in,bbllx=20pt,bblly=100pt,bburx=550pt,bbury=700pt}
\vskip -0.6 truein
\figcaption[]{The average line profile of emission from the eastern and western
sides of NGC 1232 (histogram).  The line profile is well fit by a Gaussian with a
velocity dispersion of 10.0 \kms~(thin solid line).  \label{fig:prof} }

\psfig{figure=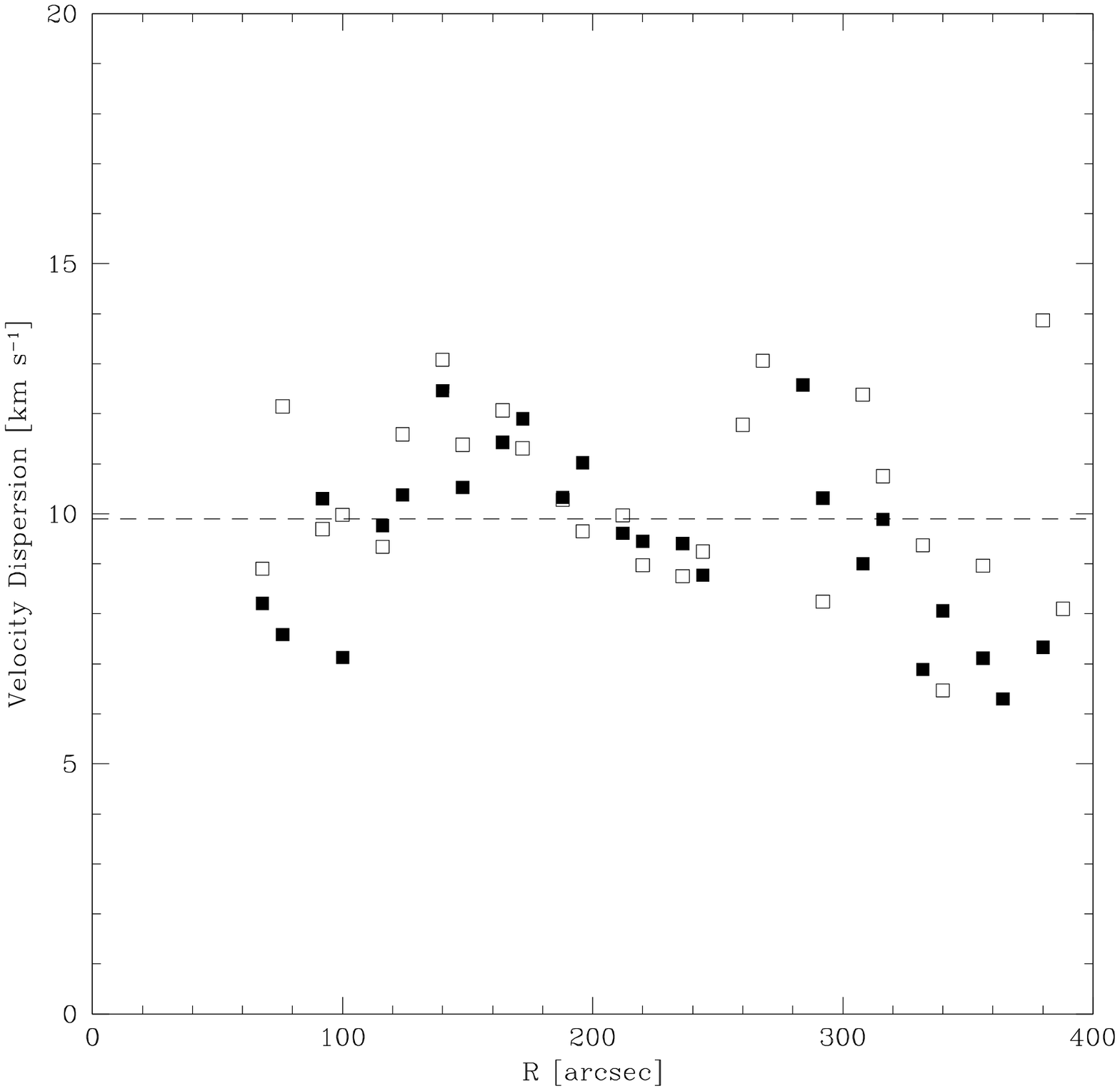,width=6.in,bbllx=20pt,bblly=100pt,bburx=550pt,bbury=700pt}
\vskip -0.3 truein
\figcaption[]{ The velocity dispersion as a function of radius.  The velocity 
dispersion was measured for spectra along two tracks parallel to the major axis
of NGC 1232 (denoted by open and filled data points).
There is no indication of a velocity dispersion fall off with radius.
The dotted line indicates the calculated average of 
all the data points, 9.9 $\pm$ 1.8 \kms.   \label{fig:disp} }

\psfig{figure=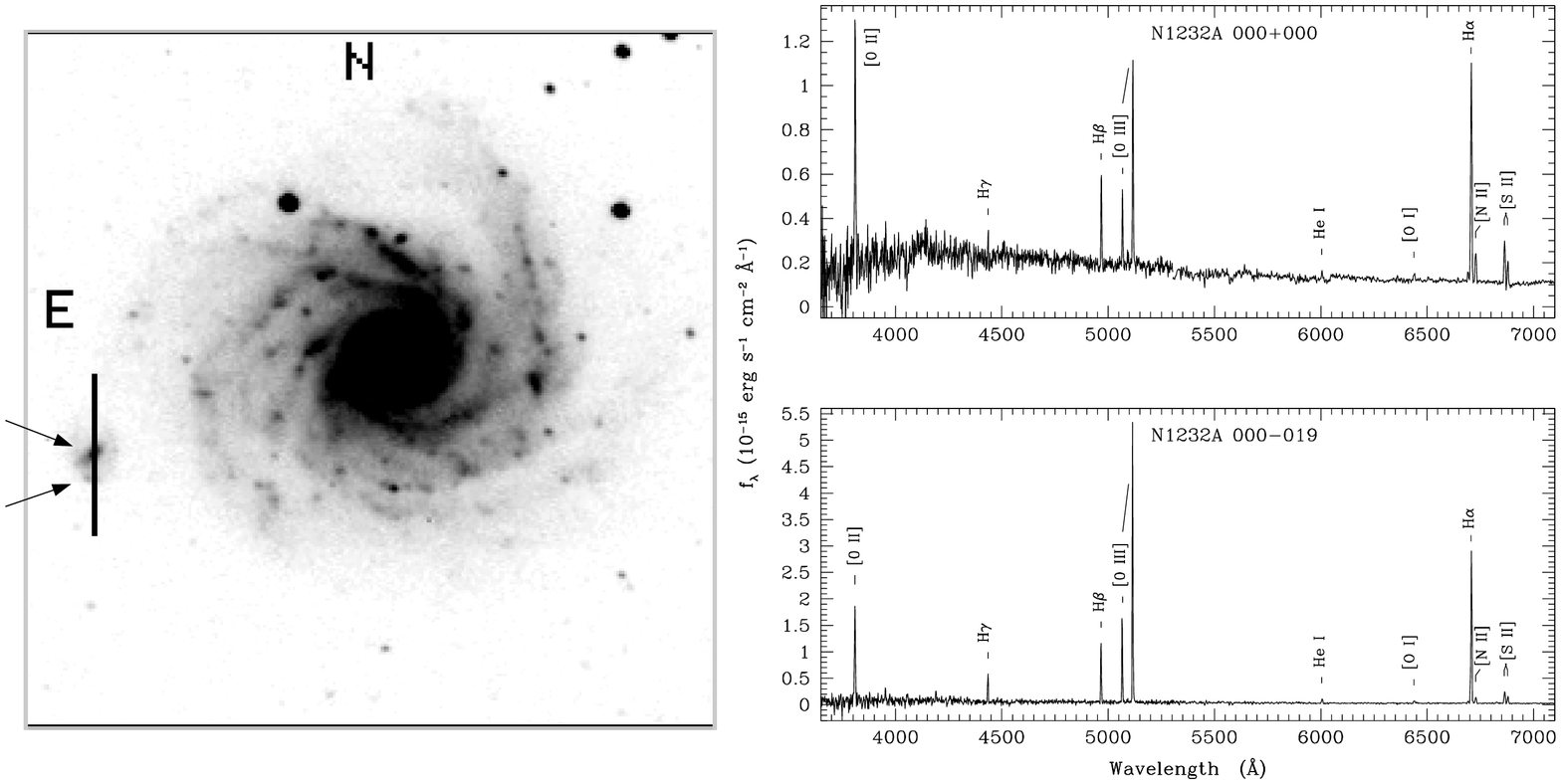,angle=270.,width=6.in,bbllx=0pt,bblly=40pt,bburx=500pt,bbury=650pt}
\figcaption[]{ (left) R--band image of NGC 1232 and NGC 1232A showing the location
of the long slit. (right) Optical spectra of the two HII regions observed in
NGC 1232A. The major lines are marked. \label{fig:n1232a} }

\begin{references}

\reference{BMRvW88} Becker, R., Mebold, U., Reif, K., \& van Woerden, H. 1988, \aap, 203, 21
\reference{BV92}   Boulanger, F., \& Viallefond, F. 1992, \aap, 266, 37
\reference{B95}    Briggs, D. 1995, PhD Thesis, New Mexico Tech
\reference{BB86}   Brinks, E., \& Bajaja, E. 1986, \aap, 169, 14
\reference{B92}    Broeils, A. H. 1992, PhD thesis,  University of Groningen
\reference{BR97}   Broeils, A. H., \& Rhee, M.--H. 1997, \aap, 324, 877
\reference{BvW94}  Broeils, A. H., \& van Woerden, H. 1994, \aaps, 107, 129
\reference{CKBvG94} Cayatte, V., Kotanyi, C., Balkowski, C., \& van Gorkom, J. H. 1994,
              \aj, 107, 1003
\reference{C87}    Condon, J. J. 1987, \apjs, 65, 485
\reference{d96}    de Jong, R. S. 1996, \aap, 313, 377
\reference{DDH87} De Robertis, M. M., Dufour, R. J., \& Hunt, R. W. 1987, \jrasc, 81, 195
\reference{RC3} de Vaucouleurs, G., de Vaucouleurs, A., Corwin, H. G., Buta, R.,
             Paturel, G., \& Fouqu\'e, P. 1991, Third Reference Catalogue
             of Bright Galaxies (Springer, New York) (RC3)
\reference{DHH90}  Dickey, J. M., Hanson, M. M., \& Helou, G. 1990, \apj, 352, 522
\reference{FT81}   Fisher, J. R. \& Tully, R. B. 1981, \apjs, 47, 139
\reference{FGW98}  Ferguson, A. M. N., Gallagher, J. S., \& Wyse, R. F. G. 1998a, \aj, 116, 673
\reference{FWGH98} Ferguson, A. M. N., Wyse, R. F. G., Gallagher, J. S., \& Hunter, 
              D. A. 1998b, \apjl, 506, L19
\reference{H98}    Haynes, M. P., Hogg, D. E., Maddalena, R. J., Roberts, M. S., \&  van Zee, L. 
               1998, \aj, 115, 62
\reference{JBH96}  Jore, K. P., Broeils, A. H., \& Haynes, M. P. 1996, \aj, 112, 438
\reference{K93}    Kamphuis, J. 1993, PhD thesis,  University of Groningen
\reference{KB92}   Kamphuis, J., \& Briggs, F. 1992, \aap, 253, 335
\reference{KH88}   Kulkarni, S. R., \& Heiles, C. 1988, in Galactic and Extragalactic
               Radio Astronomy, G.L. Verschuur and K.I. Kellermann, eds.,
               (Springer-Verlag: New York)
\reference{MRS85}  McCall, M. L., Rybski, P. M., \& Shields, G. A. 1985, \apjs, 57, 1 
\reference{M91}    McGaugh, S. S. 1991, \apj, 380, 140
\reference{AIPS}   Napier, P. J., Thompson, R. T., \& Ekers, R. D. 1983, Proc. IEEE, 71, 1295
\reference{Oke90}  Oke, J. B. 1990, \aj, 99, 1621
\reference{R82}    Reif, K., Mebold, U., Goss, W. M., van Woerden, H., \& Siegman, B. 1982, 
                       \aaps, 50, 451
\reference{RH94}   Roberts, M. S., \& Haynes, M. P. 1994, \araa, 32, 115
\reference{RDH94}  Rownd, B. K., Dickey, J. M., \& Helou, G. 1994, \aj, 108, 1638
\reference{SB99}   Sellwood, J. A., \& Balbus, S. A. 1999, \apj, 511, 660
\reference{SvK84}  Shostak, G. S., \& van der Kruit, P. C. 1984, \aap, 132, 20
\reference{SBNS89} Soifer, B. T., Boehmer, L., Neugebauer, G. \& Sanders, D. B. 1989, \aj, 98, 766
\reference{SC88}   Sparke, L. S., \& Casertano, S. 1988, \mnras, 234, 873
\reference{SSD88}  Staveley--Smith, L., \& Davies, R. D. 1988, \mnras, 231, 833
\reference{GIPSY}  van~der~Hulst, J. M., Terlouw, J. P., Begemen, K., Zwitser, W., \& 
              Roelfsema, P. R. 1992, in Astronomical Data Analysis Software and Systems, 
              ed. D. Worall, C. Biemesderfer, and J. Barnes (ASP Conference Series),  25, 131
\reference{vKS82}  van der Kruit, P. C., \& Shostak, G. S. 1982, \aap, 105, 351
\reference{vKS84}  van der Kruit, P. C., \& Shostak, G. S. 1984, \aap, 134, 258
\reference{vZ98a}  van Zee, L., Salzer, J. J., \& Haynes, M. P. 1998a, \apjl, 497, L1
\reference{vZ98b}  van Zee, L., Salzer, J. J., Haynes, M. P., O'Donoghue, A. A., \&  
              Balonek, T. J. 1998b, \aj, 116, 2805
\reference{VE92}  Vila--Costas, M. B., \& Edmunds, M. G. 1992, \mnras, 259, 121
\reference{ZKH94} Zaritsky, D., Kennicutt, R. C., \& Huchra, J. P. 1994, \apj, 420, 87  
\end{references}
\end{document}